\documentclass[preprint,prc,aps,showpacs,showkeys,groupedaddress,floatfix,natbib]{revtex4-1}
\usepackage{epsfig}
\usepackage{dcolumn}
\usepackage{bm}
\usepackage[latin5]{inputenc}
\usepackage{graphics}
\usepackage{graphicx}
\usepackage{epstopdf}
\usepackage{epsfig}
\usepackage{amssymb}
\usepackage{amsmath}
\begin{document}
\title{The comparative effect of an addition of a surface term to the Woods-Saxon potential on the thermodynamics of a nucleon}
\author{B.C. L\"{u}tf\"{u}o\u{g}lu }
\affiliation{Department of Physics, Akdeniz University, 07058 Antalya, Turkey}
\email{bclutfuoglu@akdeniz.edu.tr}
\date{\today}
\begin{abstract}
In this study, we reveal the difference between Woods-Saxon (WS) and Generalized Symmetric Woods-Saxon (GSWS) potentials in order to describe the physical properties of a nucleon, by means of solving Schr\"odinger eq. for the two potentials. The additional term squeezes the WS potential well, which leads an upward shift in the spectrum, resulting in a more realistic picture. The resulting GSWS potential does not merely accommodate extra quasi bound states, but also has modified bound state spectrum. As an application, we apply the formalism to a real problem, an $\alpha$ particle confined in Bohrium-$270$ nucleus. The thermodynamic functions Helmholtz energy, entropy, internal energy, specific heat of the system are calculated and compared for both wells. The internal energy and the specific heat capacity increase as a result of upward shift in the spectrum. The shift of the Helmholtz free energy is a direct consequence of the shift of the spectrum. The entropy decreases because of a decrement in the number of available states.
\end{abstract}
\maketitle 

\section{Introduction}

In recent years, the thermodynamic functions gained popularity in order to understand the physical properties of numerous potentials in relativistic or non-relativistic regimes. Hassanabadi \emph{et al.} studied thermodynamic properties of the three-dimensional Dirac oscillator with Aharonov-Bohm field and magnetostatic monopole potential \cite{Hassanabadi2015}. Pacheco \emph{et al.} analyzed one-dimensional Dirac oscillator in a thermal bath and they showed that its heat capacity is two times greater than that of the one-dimensional harmonic oscillator for high temperatures \cite{Pacheco2003}. Franco-Villafa\~ne \emph{et al.}  performed the first experimental study on one-dimensional Dirac oscillator \cite{Villafane2013}. Later, Pacheco \emph{et al.} also studied three-dimensional Dirac oscillator in a thermal bath \cite{Pacheco2014}. They reported that the degeneracy of energy levels and their physical implications implied that, at high temperatures, the limiting value of the specific heat is three times bigger than that of the one-dimensional case. Boumali studied the properties of the thermodynamic quantities of the relativistic harmonic oscillator using the Hurwitz zeta function. He compared his results with those obtained by a method based on the Euler-MacLaurin approach \cite{Boumali2015}. Boumali also showed that, with the concept of effective mass, the model of a two-dimensional Dirac oscillator can be used to describe the thermal properties of graphene under a uniform magnetic field, where all thermodynamic properties of graphene were calculated using the zeta function \cite{Boumali2014}. He also studied the thermodynamics of the one-dimensional Duffin-Kemmer-Petiau oscillator via the Hurwitz zeta function method \cite{Boumali2015dkp} in which study, he calculated the free energy, the total energy, the entropy, and the specific heat. Larkin \emph{et al.} have studied thermodynamics of relativistic Newton-Wigner particle in external potential field \cite{Larkin2015}. Vincze \emph{et al.} investigated nonequilibrium thermodynamic and quantum model of a damped oscillator \cite{Vincze2015}. Arda \emph{et al.} studied thermodynamic quantities such as the mean energy, Helmholtz free energy, and the specific heat with the Klein-Gordon, and Dirac equations \cite{Arda2015}. Dong \emph{et al.} studied hidden symmetries and thermodynamic properties for a harmonic oscillator plus an inverse square potential \cite{Dong2006}.

The WS potential well \cite{WoodsSaxon1954} is widely employed to model the physical systems in nuclear  \cite{WoodsSaxon1954, ZaichenkoOlkhovskii1976, PereyPerey1968, SchwierzWiedenhoverVolya, MichelNazarewiczPloszajczakBennaceur2002, MichelNazarewiczPloszajczak2004, EsbensenDavids2000,Ikhdair2013,Gautam2014, BrandanSatchler1997}, atom-molecule \cite{BrandanSatchler1997, Satchler1991}, relativistic \cite{Kennedy2002, PanellaBiondini2010, AydogduArda2012, GuoSheng2005, GuoZheng2002, RojasVillalba2005, HassanabadiMaghsoodi2013, YazarlooMehraban2016, Chargui2016} and  non-relativistic \cite{PahlavaniAlavi2012, CostaPrudente1999, Fluge1994, Saha2011, Feizi2011, NiknamRajabi2016} physics problems.

To describe the energy barrier at the surface of atomic nucleus that nucleons are exposed, various type of additional terms to WS potential are proposed to produce GSWS potentials. Such  potential wells can be used to model any system, in which a particle is trapped in a finite space, as well as the effects, such as non-zero $l$, spin-orbit coupling \cite{CandemirBayrak2014, BayrakSahin2015, BayrakAciksoz2015, LutfuogluAkdeniz2016, LiendoCastro2016, BerkdemirBerkdemir2005, BadalovAhmado2009, GonulKoksal2007, KouraYamada2000, CapakPetrellis2015, CapakGonul2016, IkotAkpan2012, IkhdairFalayeHamzavi2013, surface1, surface2, surface3, surface4, surface5}.

Our main motivation in this work is to compare physical consequences of the two potentials in context of quantum mechanics and statistical thermodynamics. We consider the physical properties of $\alpha$ particle as an application, to reinforce the formal treatment of the two potentials for Bh-$270$ nucleus.

In section 2, we interpret the forms of the WS and GSWS potentials, and corresponding energy eigenvalues, for a massive non-relativistic confined particle, using the formalism proposed by \cite{LutfuogluAkdeniz2016}. In section 3, we give a brief summary on the thermodynamic functions, that are calculated in the following section for the two potentials. In section 4, the energy spectra of $\alpha$ particle in Bh-$270$ nucleus for the two potentials are presented as an application of the formalism presented, upon which, the thermodynamic functions of the system are plotted and discussed in terms of the parameters of the problem. In section 5, the conclusion is given.

\section{The Model}\label{themodel}
The WS potential well in one dimension is described by
\begin{eqnarray}\label{wsp}
  V(x)&=&-\theta{(-x)}\frac{V_0}{1+e^{-a(x+L)}}- \theta{(x)}\frac{V_0}{1+e^{a(x-L)}}, \label{ws}
 \end{eqnarray}
where $\theta{(\pm x)}$ are the Heaviside step functions, $a$ is the reciprocal of the diffusion coefficient, $L$ measures the size of the nucleus, $V_0$ is the depth of the potential, given by \cite{PereyPerey1968}
\begin{eqnarray}
  V_0 &=& 40.5+ 0.13 A, \label{potentialdepth}
\end{eqnarray}
where $A$ is the atomic number of the nucleus.

According to the assumption that a nucleon suffers a potential barrier when near the surface of its nucleus or being emitted to outside, the WS potential is considered inadequate to explain the dynamics of this type of problems. In order to take the surface effect into account, an additional term to the WS potential is widely used \cite{BayrakAciksoz2015, BayrakSahin2015, LutfuogluAkdeniz2016}. The WS potential combined with the additional terms are called GSWS potential.
\begin{eqnarray}\label{gws}
  V(x)&=&\theta{(-x)}\Bigg[-\frac{V_0}{1+e^{-a(x+L)}}+\frac{W_0 e^{-a(x+L)}}{\big(1+e^{-a(x+L)}\big)^2}\Bigg] \nonumber \\
  &+& \theta{(x)}\Bigg[-\frac{V_0}{1+e^{a(x-L)}}+\frac{W_0 e^{a(x-L)}}{\big(1+e^{a(x-L)}\big)^2}\Bigg], \label{gsws}
 \end{eqnarray}
here the second terms in the brackets correspond to the energy barrier that nucleon faces at the surface, which is taken as linearly proportional to the spatial derivative of the first term multiplied by the nuclear size. A unitless proportionality multiplier, hereby $\rho$, which is implicitly included in $W_0$, can be calculated via conservation laws.

Because of the symmetry of the potential, even $E_n^{e}$ and odd $E_n^{o}$ energy eigenvalues arise, which are studied extensively in the reference \cite{LutfuogluAkdeniz2016} and evaluated to be
\begin{eqnarray}
E_n^{e}&=&-V_0+\frac{\hbar^2}{2 m L^2}\Bigg|\arctan \frac{(N_1-N_2)}{i(N_1+N_2)}\pm n'\pi\Bigg|^2,  \\
E_n^{o}&=&-V_0+\frac{\hbar^2}{2 m L^2}\Bigg|\arctan \frac{(N_1+N_2)}{i(N_1-N_2)}\pm n'\pi\Bigg|^2,
\end{eqnarray}
here $n'$ are integers, whereas $n$ stands for the number of nodes, the roots of the wave functions. $N_1$ and $N_2$ are complex numbers
\begin{eqnarray}
  N_1 &=& \frac{\Gamma(c_1)\Gamma(c_1-a_1-b_1)}{\Gamma(c_1-a_1)\Gamma(c_1-b_1)}, \\
  N_2 &=& \frac{\Gamma(c_1)\Gamma(a_1+b_1-c_1)}{\Gamma(a_1)\Gamma(b_1)},
\end{eqnarray}
and implicitly dependent on the energy eigenvalues via the coefficients $a_1$, $b_1$ and $c_1$
\begin{eqnarray}
  a_1 &=& \mu+\theta+\nu, \\
  b_1 &=& 1+\mu-\theta+\nu, \\
  c_1 &=& 1+2\mu,
\end{eqnarray}
here
\begin{eqnarray}
  \mu       &=& \sqrt{-\frac{2 m E_n}{a^2 \hbar^2}}, \\
  \nu       &=& \sqrt{-\frac{2 m (E_n+V_0)}{a^2 \hbar^2}}, \\
  \theta    &=& \frac{1}{2}\mp\sqrt{\frac{1}{4}-\frac{2mW_0}{a^2\hbar^2}}.
\end{eqnarray}
When $W_0=0$, $N_1$ and $N_2$ remain unchanged because of the symmetry in the multiplication of the Gamma functions under the possible values of $\theta$, which are either $0$ or $1$. Moreover, since the whole energy spectrum is negative, $\mu$ is real, the ordinary solutions for the WS potential are obtained. When $W_0$ is between $0$ and $V_0$, the WS potential well is slightly modified because of being narrower, but not yet giving rise to positive energy eigenvalues. When $W_0$ exceeds $V_0$, the barrier starts to grow and the well keeps narrowing, this alters the energy spectrum, including an extension to positive values. The positive energies are the reason for complex values of $\mu$, which are responsible for tunnelling in some nuclei. These states are called quasi-bound states.

The energy spectrum of a nucleon under GSWS potential is composed of energy eigenvalues, satisfying
\begin{eqnarray}
  -V_0<E_n<V_0\frac{(1-\rho aL)^2}{4\rho aL}
\end{eqnarray}
then, $\nu$ can take only imaginary values for the entire scope of the spectrum.
\section{Thermodynamics of a System}\label{thermosection}

Using the energy eigenvalues $E_n$, the partition function of the system is given by

\begin{eqnarray}
  Z(\beta) &=& \sum_{n} e^{-\beta E_n}, \label{partitionfunction}
\end{eqnarray}
where $\beta$ is defined by
\begin{eqnarray}
  \beta &=& \frac{1}{k_B T}.
\end{eqnarray}
and $k_B$ stands for the Boltzmann constant, $T$ is the temperature in the unit of Kelvin. The Helmholtz free energy of the system can be calculated using the equation
\begin{eqnarray}
      F(T) &\equiv& -k_B T\ln Z(\beta).
    \end{eqnarray}
The entropy of the system is given by,
  \begin{eqnarray}
      S(T) &=& -\frac{\partial}{\partial T}F(T).
    \end{eqnarray}
The internal energy $U(T)$ is the expectation value of the energy of the system
\begin{eqnarray}
      U(T) &=& -\frac{\partial}{\partial \beta}\ln Z(\beta).
    \end{eqnarray}
Then, the isochoric specific heat capacity $C_v(T)$  is defined by
\begin{eqnarray}
      C_v(T) &\equiv& \frac{\partial}{\partial T}U(T).
    \end{eqnarray}

\section{An application of the formalism for Bh-$270$ nucleus}\label{application}

\begin{figure}[htb]
\centerline{\includegraphics[width=12.5cm]{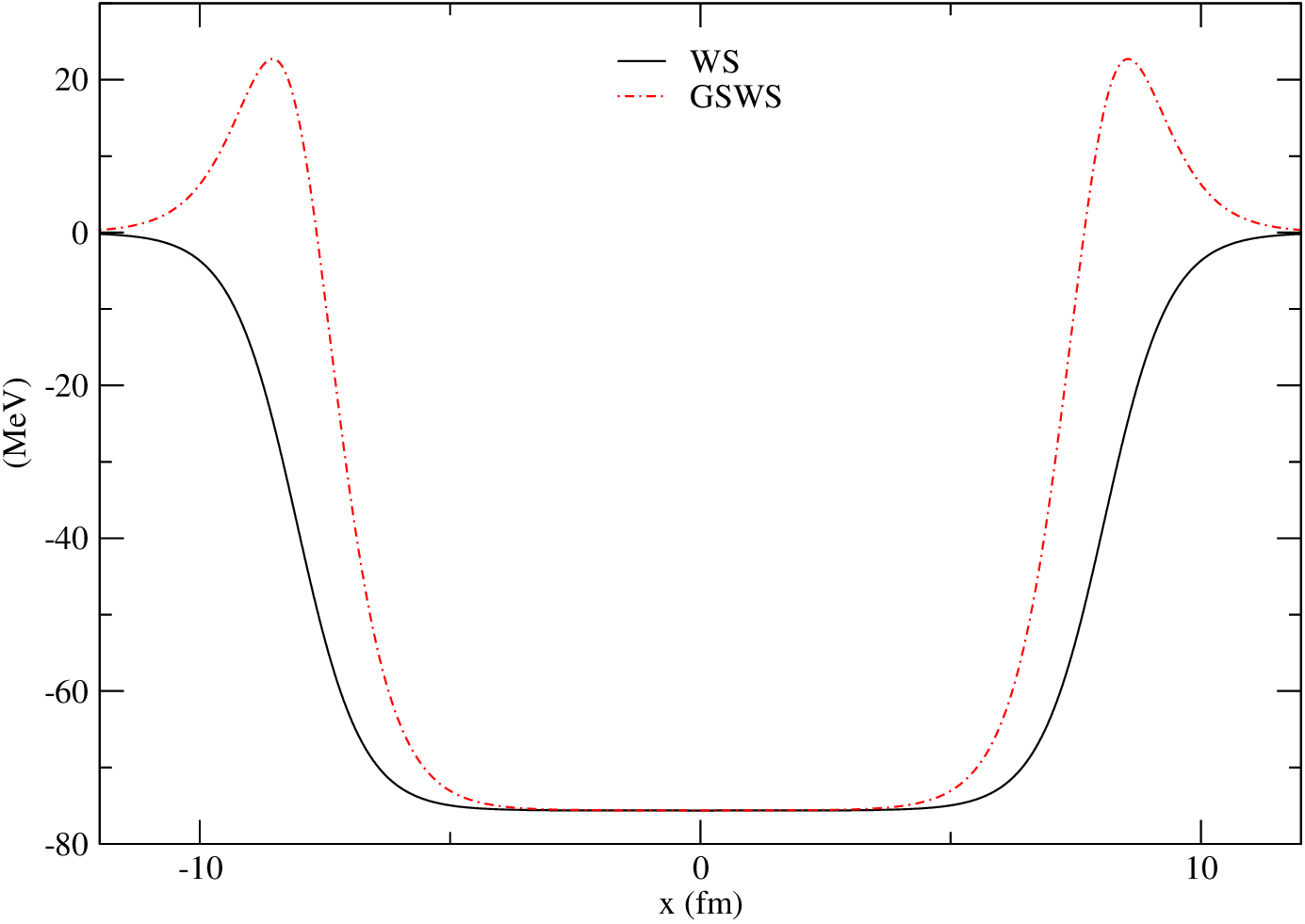}}
\caption{  The WS and GSWS potential well for an $\alpha$ particle in a Bh-$270$ nucleus. }
\label{fig1}
\end{figure}

In this section we present the thermodynamic treatment of an $\alpha$ particle within Bh-$270$ nucleus as an application of the formalism described in previous sections, in order to investigate the effects of the surface term addition to the WS potential. For this nucleus, in ref \cite{PereyPerey1968} the inverse diffusion parameter  is given as $a=1.538 fm^{-1}$, while the  radius is evaluated to be $L=8.068 fm$. Substituting the atomic number $A=270$ into (\ref{potentialdepth}), we have $V_0=75.617 MeV$ and then $W_0=215.523 MeV$.
The corresponding WS and GSWS potentials are shown in Fig.~\ref{fig1}.
\\
\\

The calculated energy spectra of an $\alpha$ particle with mass $m=3727.379 MeV/c^2$ in the nucleus are tabulated in  Table $1$ and Table $2$  for WS and GSWS, respectively. Purely real bound state energy eigenvalues in both spectra imply infinite time constants, which mean zero decay probability for the nucleon from the nucleus. Whereas, the quasi-bound states in the GSWS spectrum have a complex form with finite time constants, which are responsible for the decay probability \cite{Gamow1928, Siegert1939}.

\begin{table}[pt]
\caption{The energy spectrum of the $\alpha$ particle within Bh-$270$ nucleus under WS potential well assumption.}
\begin{tabular}{@{}cccccccc@{}} \hline
$n$ &$E_n$& $n$ &$E_n$& $n$ &$E_n$& $n$ &$E_n$ \\ \label{table:energyspectrumWS}
& (MeV)& & (MeV)&&(MeV)&& (MeV) \\ \hline
$0$&$-75.283$ &      $6$&$-63.011$&        $12$&$-39.441$    &$18$&$-10.838$\\
$1$&$-74.336$ &      $7$&$-59.712$&        $13$&$-34.806$    &$19$&$-6.452$\\
$2$&$-72.873$ &      $8$&$-56.130$&        $14$&$-30.050$    &$20$&$-2.654$\\
$3$&$-70.967$ &      $9$&$-52.289$&        $15$&$-25.213$    &$21$&$-0.171$\\
$4$&$-68.665$ &      $10$&$-48.210$&       $16$&$-20.347$    &$$  &$$\\
$5$&$-66.004$ &      $11$&$-43.919$&       $17$&$-15.522$    &$$  &$$\\ \hline
\end{tabular}
\end{table}

\begin{table}[pt]
\caption{The energy spectrum of the $\alpha$ particle within Bh-$270$ nucleus under GSWS potential well assumption. The rightmost column tabulates the quasi-bound energy levels.}
\begin{tabular}{@{}cccccccc@{}} \hline
$n$ &$E_n$ & $n$ &$E_n$& $n$ &$E_n$& $n$ &$E_n$\\
& (MeV)& & (MeV)&&(MeV)&& (MeV) \\ \hline
$0$&$-75.166$&$6$&$-59.531$&$12$&$-29.605$&$17$&$2.263- 0.537 \times 10^{-3} i$\\
$1$&$-73.915$&$7$&$-55.373$&$13$&$-23.607$&$18$&$8.929 - 0.146 \times 10^{-1} i$\\
$2$&$-72.022$&$8$&$-50.855$&$14$&$-17.386$&$19$&$15.439 - 0.133 i$\\
$3$&$-69.585$&$9$&$-45.998$&$15$&$-10.971$&$20$&$21.688 - 0.650 i$\\
$4$&$-66.666$&$10$&$-40.823$&$16$&$-4.402$&$$&$$\\
$5$&$-63.304$&$11$&$-35.351$&$$&$$&$$&$$\\
\hline\label{table:energy_spectrum GSWS}
\end{tabular}

\end{table}

\begin{figure}[htb]
\centerline{\includegraphics[width=12.5cm]{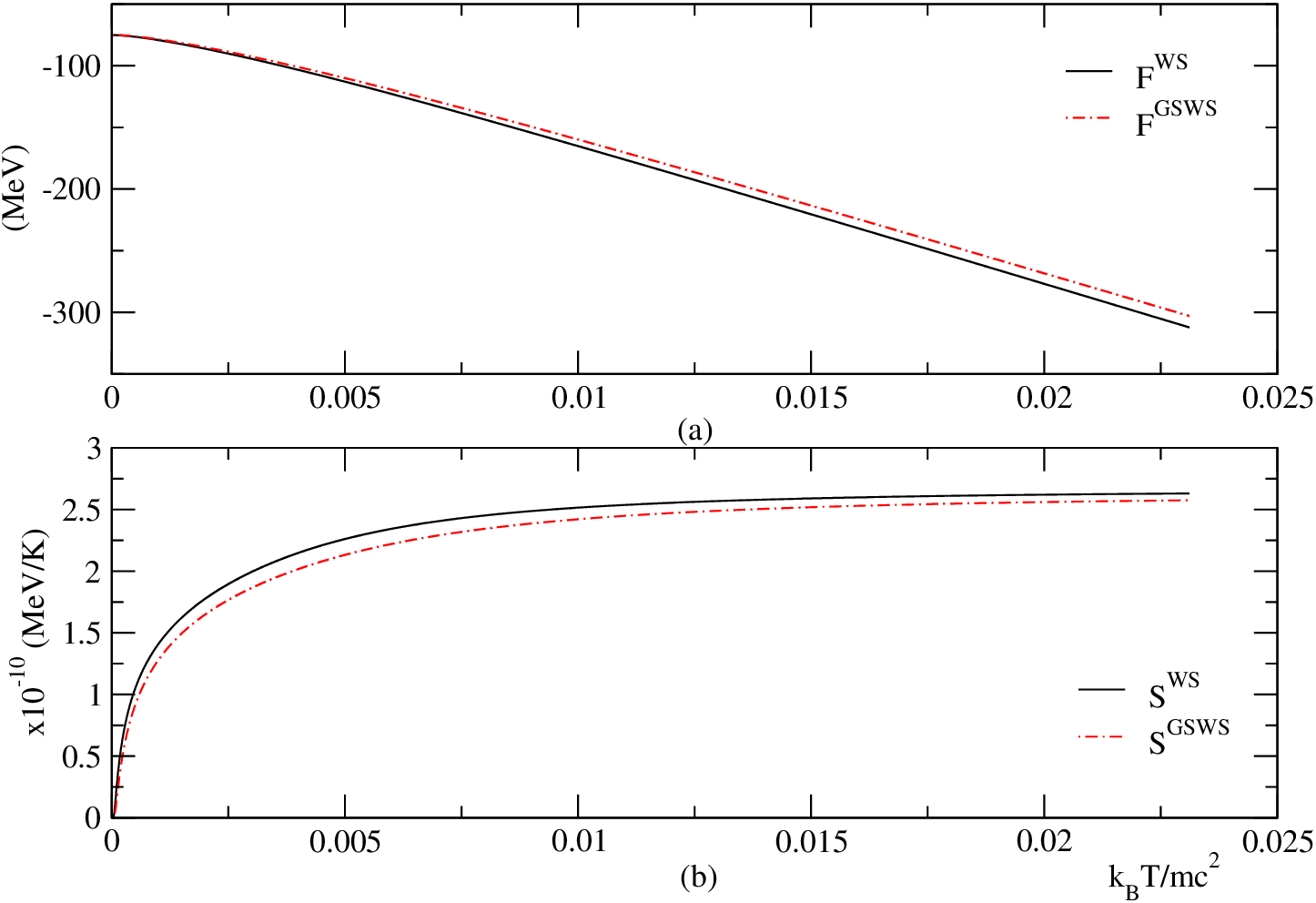}}
\caption{Helmholtz energy $F(T)$ (a), entropy $S(T)$ (b), as functions of reduced temperature. }
\label{Bohrium_270_Helmholtz_Free_Energy_and_Entropy_WS_vs_GWSW}
\end{figure}

Using the partition function given in (\ref{partitionfunction}), the Helmholtz and the entropy functions versus reduced temperature curves of the system, corresponding to the cases of GS and GSWS potentials are presented in Fig.~\ref{Bohrium_270_Helmholtz_Free_Energy_and_Entropy_WS_vs_GWSW} (a) and (b), respectively. The entropy in both cases start from zero, being in agreement with the third law of thermodynamics. The saturation values of the entropy are $2.66\times 10^{-10} MeV/K$ and $2.62\times 10^{-10} MeV/K$ for WS and GSWS potentials, respectively. Surprisingly the addition of the the surface term does not lead to an increase in the number of available states, contrarily, it results in a decrease in the number of available states from $22$ to $21$, accompanied by an upward shift in the energy spectrum. This is a consequence of the upper shift of the energy spectrum by \textit{squeezing} the well with the addition. The increase of the Helmholtz free energy is due to the upward shift in the energy spectrum.
\\

\begin{figure}[htb]
\centerline{\includegraphics[width=12.5cm]{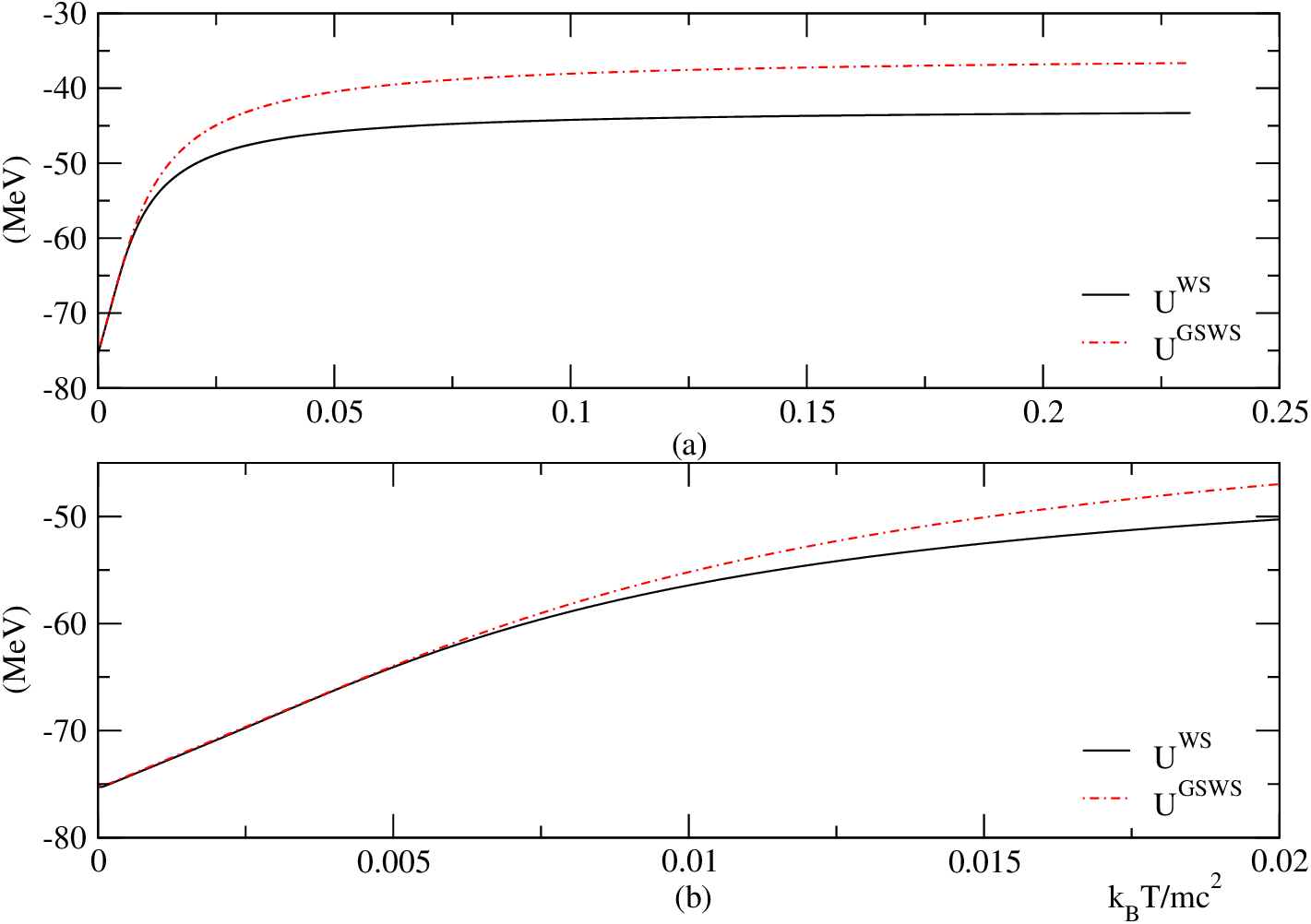}}
\caption{  Internal energy $U(T)$ as functions of reduced temperature (a), the initial behavior (b). }
\label{Bohrium_270_Internal_Energy_WS_vs_GWSW}
\end{figure}

The addition of the surface term leads to increase in the internal energy as observed in Fig.~\ref{Bohrium_270_Internal_Energy_WS_vs_GWSW}, since it is the expectation value of the energy eigenvalues $E_n$. The internal energies initiate at the values $-75.283 MeV$ and $-75.166 MeV$ at $0 K$ for the WS and GSWS potentials, which are the lowest energy eigenvalues in the spectra, respectively. These internal energies are not distinguishable to the naked eye until the reduced temperature of about $0.007$, at which the difference broadens, as observed in Fig.~\ref{Bohrium_270_Internal_Energy_WS_vs_GWSW} (b). The limiting values of the internal energies of the two cases goes to the mean values of the $-42.586 MeV$ and $-35.535 MeV$ for the spectra of WS and GSWS, respectively, as the reduced temperature goes to infinity.
\\

\begin{figure}[htb]
\centerline{\includegraphics[width=12.5cm]{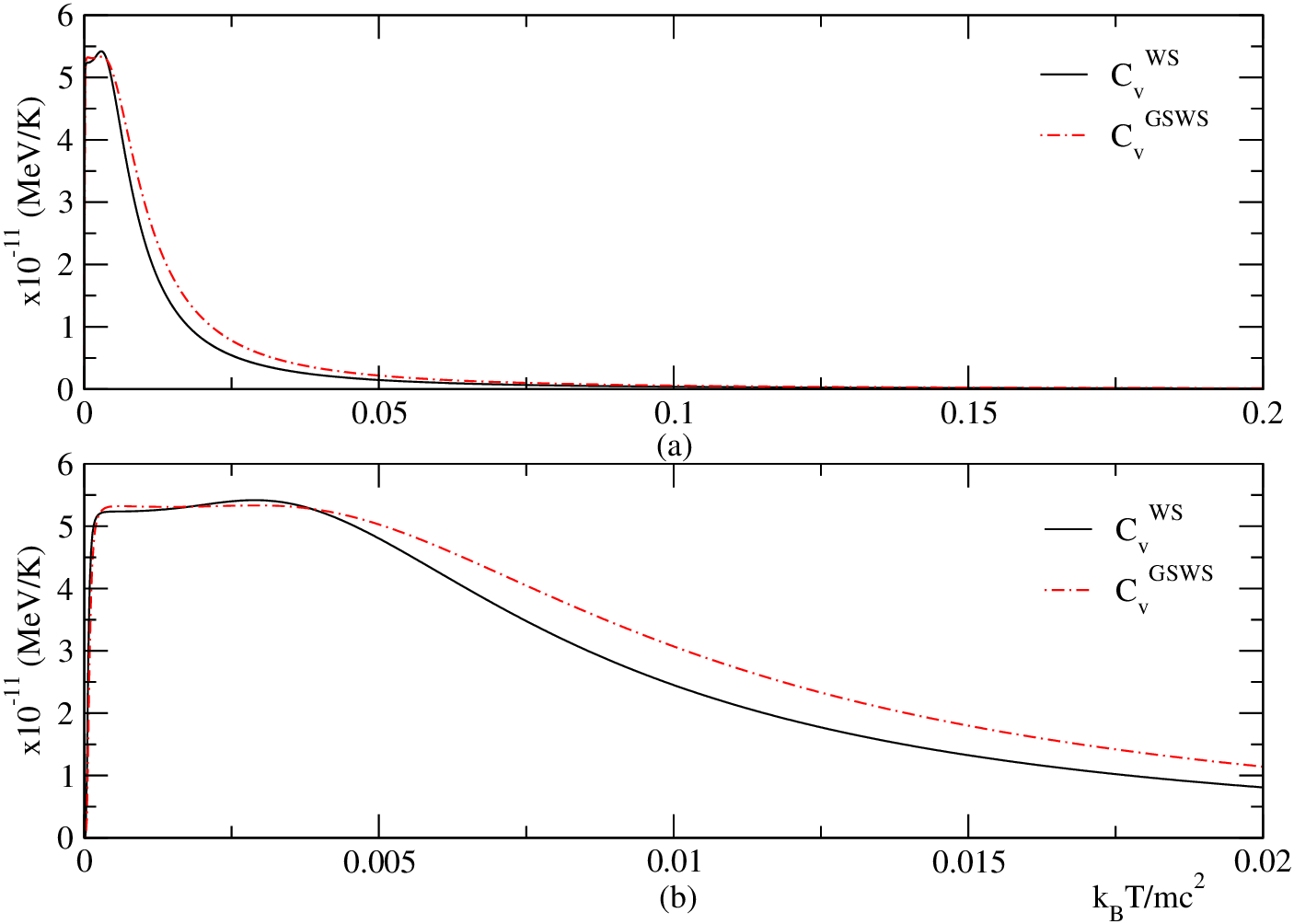}}
\caption{  Specific heat $C_v(T)$ as functions of reduced temperature (a), the initial behavior (b). }
\label{Bohrium_270_Specific_Heat_and_Inset_WS_vs_GWSW}
\end{figure}

In Fig.~\ref{Bohrium_270_Specific_Heat_and_Inset_WS_vs_GWSW}, the specific heat $C_v(T)$ versus reduced temperature curves are demonstrated. The steep linear initial increase is a consequence of the initial convex behavior of the internal energies, which is a common characteristics of the two cases. After that, $C_v(T)$ remains constant, followed by a decay to zero, in the whole scale of the reduced temperature. The specific heat function for GSWS potential has higher values during this decay, which verifies that the internal energy saturates to a higher value at higher reduced temperature.
\\

\section{Conclusion}\label{conclusion}

In this study, we analyze the effect of the additional term, representing the surface effect of a nucleus, to WS potential well. We formally discuss how the additional term modifies the whole non-relativistic energy spectrum by squeezing the well, resulting in an upward shift of the spectrum. GSWS potential does not merely accommodate extra quasi bound states, but also has modified bound state spectrum. As an application of the formal treatment, we consider $\alpha$ particle inside Bh-$270$ nucleus, modeled with both WS and GSWS potential wells. The thermodynamic functions Helmholtz free energy, entropy, internal energy, specific heat are calculated in both approaches and compared. The internal energy and the specific heat capacity increase, as a result of upward shift in the spectrum. The shift of the Helmholtz free energy is a direct consequence
of the shift of the spectrum. The entropy decreases due to the decrement in the number of available states, which arises as a result of narrowing the well with the additional term. It is concluded that GSWS potential is more realistic to describe the physical properties of $\alpha$ particle within Bh-$270$ nucleus.

\section*{Acknowledgments}
This work was partially supported by the Turkish Science and Research Council (T\"{U}B\.{I}TAK) and Akdeniz University.

The author would like to thank to Dr. M. Erdogan for all scientific discussions and the preparation of this manuscript. The author also thanks Prof. I. Boztosun and E. Pehlivan for sharing their valuable comments on the manuscript.

\vspace{-1mm}
\centerline{\rule{80mm}{0.1pt}}

\end{document}